\documentclass[10pt,a4paper]{article}
\usepackage[T1]{fontenc}
\usepackage{lmodern}
\usepackage[margin=2.3cm]{geometry}
\usepackage{amsmath,amssymb,booktabs,tabularx,array}
\usepackage{graphicx,xcolor,microtype}
\usepackage{enumitem}
\usepackage{tikz,pgfplots}
\usetikzlibrary{arrows.meta,positioning}
\pgfplotsset{compat=1.18}
\usepackage[numbers,sort&compress]{natbib}
\usepackage{xurl}
\usepackage[colorlinks=true,linkcolor=blue!50!black,citecolor=green!35!black,urlcolor=blue!50!black]{hyperref}
\usepackage{caption}
\setlist{nosep,leftmargin=1.5em}
\newcolumntype{Y}{>{\raggedright\arraybackslash}X}

\definecolor{teal}{HTML}{087F8C}
\definecolor{brick}{HTML}{B33C4A}
\definecolor{ink}{HTML}{303A46}
\definecolor{ochre}{HTML}{A97718}

\title{\textbf{Selecting and Composing Agent Design Patterns:}\\[4pt]
Industrial Case Evidence and Controlled\\Reference-System Experiments}
\author{Jia Huang\textsuperscript{1}\quad Bingsheng Ru\textsuperscript{2}\quad Bo Liang\textsuperscript{3}\\[3pt]
Yuke Xiong\textsuperscript{4}\quad Haiyan Fan\textsuperscript{5}\quad Joey Tianyi Zhou\textsuperscript{1,6}\\[6pt]
\small\textsuperscript{1}Agency for Science, Technology and Research (A*STAR), Singapore\\
\small\textsuperscript{2}Tencent Research Institute, China\\
\small\textsuperscript{3}Shanghai Dongfang Yiteng Technology Co., Ltd., China\\
\small\textsuperscript{4}Chengdu Xuansu Technology, China\\
\small\textsuperscript{5}Wanlian Yida Holding Group Co., Ltd., China\\
\small\textsuperscript{6}Centre for Frontier AI Research (CFAR), A*STAR, Singapore\\[3pt]
\footnotesize\texttt{huang\_jia@a-star.edu.sg}\quad\texttt{robinru@tencent.com}\quad\texttt{lb@eteng.cn}\\
\footnotesize\texttt{being\_xiong@outlook.com}\quad\texttt{hy\_fan@yeah.net}\quad\texttt{joey\_zhou@a-star.edu.sg}}
\date{19 October 2026}

\begin{document}
\maketitle
\begin{abstract}
Agent pattern catalogs describe architectural alternatives, but engineers also need evidence for selecting and composing them. A recurring industrial difficulty is that an apparently correct step can lose its meaning during execution: a valid identifier refers to the wrong record, a successful publication returns an unusable map, or a repeated event applies an effect twice. We extend a cognitive-function--execution-topology framework with an evidence-driven selection procedure and explicit composition contracts. The study combines two attributed industrial case accounts, a contributor's collaboration design, seven practitioner workshops, and executable fault-injection experiments.

Three local reference systems operationalize the reported mechanisms using SQLite state, shared artifact files, and a file-backed map-publication surrogate. A $2\times2$ ablation separates function-specific contracts from execution controls across 36 scenario families, with ten input variants per family. Under equal family weighting, composition correctly completes or safely refuses 11 of 12 families in each system, compared with 2, 2, and 1 for the simple baselines. These are constructed challenge-suite results, not production success rates. Recovery without consumer-side verification increases false-success cases in the map surrogate. All composed systems still fail a boundary case in which trusted evidence is itself wrong. A separate audit of a frozen six-arm language-model pilot reproduces its reported scores but confirms an evaluator-dependent ranking reversal.

The contribution is a traceable connection from industrial failure to pattern selection, executable treatment, external acceptance, and revision. The selection step is demonstrated on these cases rather than evaluated against an alternative procedure, and catalog membership is governed by separately stated admission criteria rather than by these experiments. The results support specific composition mechanisms in bounded reference systems. Whether the framework improves engineers' design decisions over an equally informative conventional method remains a separate, explicitly specified research question.
\end{abstract}

\section{Introduction}
An execution agent selects the right API, receives a valid identifier, and then supplies a different identifier to the next call. A GIS pipeline publishes a layer successfully, yet the consumer receives an XML exception or a blank image. Two cooperating agents process the same event twice after a worker restarts. These failures occur at different boundaries, but they share an architectural difficulty: information that was sufficient to advance one step was insufficient to justify the next step or the final claim of completion.

The industrial material underlying this study describes concrete responses. In Bo Liang's execution-agent case, exact business values move from conversational history to program-managed state, while a task graph records dependencies and resumption points. In Yuke Xiong's GIS case, tested processing decisions become versioned pipelines, and actual map requests establish acceptance. Haiyan Fan's collaboration proposal connects shared artifact references with event delivery, access control, and lifecycle management. These accounts provide empirical evidence about engineering problems and design reasoning. They do not all provide the same kind of evidence: the first two describe implemented systems and encountered failures, while the third specifies an architecture whose deployed performance has not yet been measured in the available record.

Our earlier framework organizes agent patterns along cognitive function and execution topology~\cite{huangzhou}. A function describes the responsibility being fulfilled, such as remembering a fact, choosing an action, or evaluating an outcome. A topology describes the control structure through which that work is performed. This distinction helps expose incomplete designs. Preserving an identifier does not determine when its consuming action may run. A retry loop does not establish whether the repeated action is correct. Selection must connect both questions to an observable engineering requirement.

The central proposition investigated here is conditional: \emph{when a failure crosses a function boundary and an execution boundary, a composition that preserves the required information and enforces the required transition can address failures that either component leaves unresolved.} This proposition predicts interaction in some settings and merely additive effects in others. It also predicts no benefit, or added overhead, when the relevant failure is absent. It does not claim that every matrix cell should be used or that more agents improve a system.

We distinguish three questions throughout the paper, and separate all of them from a fourth, prior question---whether a pattern belongs in the catalog at all---which Section~\ref{sec:admission} answers with stated criteria rather than with experiments:
\begin{enumerate}
\item \textbf{Engineering relevance:} which observed industrial failures and constraints motivate a pattern or composition?
\item \textbf{Mechanism effectiveness:} what changes when an implemented contract, execution control, or their composition is enabled under the same task and injected fault?
\item \textbf{Design-method effectiveness:} do engineers using the two-axis selection method build better systems than engineers given comparable information in another form?
\end{enumerate}
The industrial accounts inform the first question. The new reference-system experiments and historical pilot audit address bounded aspects of the second. This manuscript does not report a completed human study of the third. Conflating these questions would turn a correct implementation of a known control into unsupported evidence for the causal superiority of a design method.

The paper contributes explicit admission and retirement criteria for the catalog, a case-grounded selection procedure together with its worked application, a contract for testing composition, three executable reference systems with full ablations and artifact readback, and an evidence-maintenance process that accommodates negative results and new contributor cases. The historical pilot is retained because its ranking reversal changes how architectural gains should be measured. Together, the studies provide an empirical foundation for further industrial replication without requiring an uninformative claim that the entire catalog has been universally validated.

\section{Related Work and Research Position}
Design patterns have long provided reusable descriptions of recurring software structures~\cite{gamma}. Their empirical benefits have nevertheless required task-specific investigation. Prechelt et al. compared pattern-based designs with simpler alternatives in maintenance tasks performed by professional engineers and found both favorable and unfavorable outcomes~\cite{prechelt}. A later replication examined when particular patterns help and how their use affects maintainability~\cite{vokac}. We adopt the same principle: the relevant comparison is an executable alternative under stated requirements, including conditions in which the simpler design wins.

Industrial case research retains context that controlled benchmarks remove: why a team changed its architecture, which constraints it could not change, and how an observed failure affected the decision. Case-study guidance emphasizes an explicit unit of analysis and a chain of evidence~\cite{runeson}. Our units are reported design decisions and bounded reference runs. We do not treat workshop participants as a representative industry survey, multiple documents about one project as independent deployments, or a contributor's proposal as a measured production intervention.

Agent evaluations have moved beyond final-text quality. AgentBench studies interactive behavior~\cite{agentbench}; $\tau$-bench evaluates final database states and repeated success under domain policies~\cite{taubench}. AgentDojo separates useful task execution from security failures and supports extensible adversarial cases~\cite{agentdojo}. Procedure-Aware Evaluation identifies outcomes that appear successful while violating required procedures~\cite{pae}. Our consumer acceptance and false-success measurements are aligned with this work. The distinction is the treatment unit: we compare explicit architectural components while holding the task proposal and environment condition fixed.

Industry guidance also distinguishes workflow topologies and recommends adding complexity when measurements justify it~\cite{anthropic}. The two-axis framework supplies an additional functional coordinate, not an alternative to transactions, state machines, or established workflow engineering. Recent guidelines for empirical studies involving language models emphasize model versions, system configuration, suitable baselines, traces, human validation, and disclosure of AI use~\cite{baltes}. We apply those requirements to the frozen model pilot and identify the missing metadata. The new experiments make no model calls: they isolate execution behavior after a proposal has been produced.

\section{Industrial Evidence and Study Scope}
\label{sec:cases}
\subsection{Sources and synthesis}
The evidence corpus comprises seven ADPS module workshops held on 5, 6, 12, 13, 18, 25, and 26 August 2026, two industrial case reports, a collaboration design contribution received in September, and the frozen RegTech experiment~\cite{workshops,liangcase,xiongcase,fanproposal}. The reasoning workshop on 26 August adds material that was not included in the earlier six-workshop draft, particularly scenario-specific benchmarks, migration of reasoning controls, and external acceptance of deterministic pipelines~\cite{reasoningworkshop}.

This is a purposive engineering synthesis. Source records were read to identify the business object, failure, architectural response, acceptance evidence, and remaining uncertainty. Public reports were checked against the local case material; the reasoning workshop's relevant passages were also checked against the transcript. The synthesis has not undergone independent qualitative coding or member checking for this manuscript. We therefore report neither inter-rater agreement nor frequency estimates. A statement repeated in a workshop, a book chapter, and a case report remains one linked account until independent evidence establishes otherwise.

\begin{table}[t]
\centering\small
\caption{Source types and the claims they support. Existing field material and newly executed reference experiments retain separate provenance.}
\label{tab:evidence}
\begin{tabularx}{\textwidth}{@{}p{2.5cm}YY@{}}
\toprule
Source & Available evidence & Role in this study\\
\midrule
Dongfang Yiteng & Contributor account of prototype failures, state architecture, task dependencies, and recovery design & Industrial motivation for exact parameter binding and durable execution\\
Xuansu GIS & Failure accounts, pipeline and lifecycle descriptions, console and map-output artifacts & Industrial motivation for consumer acceptance and bounded recovery\\
Shared artifacts & Fan's two design notes and interaction diagram; no supplied deployment measurements & Attributed design input for event and artifact experiments\\
Seven workshops & Dated discussion records, transcripts, questions, disagreements, and proposed revisions & Requirements for measurement and selection\\
RegTech pilot & Six frozen model outputs, one source bundle, 12 checklist items, original scorer & Historical outcome and evaluator audit\\
Reference study & Executed SQLite/file systems, injected faults, per-run traces and artifact readback & Controlled evidence for the implemented mechanisms\\
\bottomrule
\end{tabularx}
\end{table}

\subsection{Business execution: exact state and permitted order}
Liang's case concerns an agent for configuring payroll-related workflows in an existing SaaS environment~\cite{liangcase}. The prototype could discover tools and select plausible actions. The harder problem appeared across calls: a returned business identifier could be rewritten, misbound, or consumed after a skipped dependency. A schema-valid identifier was insufficient because its provenance, rather than its syntax, determined the object that would change.

The reported response separated three kinds of state. Program state retained exact values and their producing calls. Task state retained dependency and completion information. Conversational state retained user intent and explanatory context. Once the intended steps became known, a task graph determined which action could execute. Sensitive steps retained a persistent resumption point through approval. These changes explain an engineering rationale for composing Action and Memory responsibilities with ordered execution and control gates.

The available report does not give a complete before-and-after run population or a public identifier-error rate. We preserve the case as an attributed industrial account and reconstruct a smaller executable experiment. The reconstruction updates an integer-valued record in SQLite after taking a snapshot. It tests exact target binding, dependency order, repeated delivery, and interruption around a committed effect. It does not reproduce the company's full payroll platform, tenant isolation, payment integration, or human approval interface.

\subsection{GIS publication: acceptance from the consuming side}
Xiong's case follows a pipeline that validates GIS inputs, processes data, generates styling, publishes a service, creates a viewer, and verifies the result~\cite{xiongcase}. Two reported failures make the acceptance problem concrete. A layer-group configuration could lead a GetMap request to return \texttt{LayerNotDefined} inside a ServiceException despite HTTP 200. A zero metatile setting could produce a GetTile error. The case also describes coordinate-related failures that only become apparent when consuming the published map.

The team's response places explicit checks after publication. Request status, content type, exception bodies, and meaningful image output provide different pieces of evidence. Tested pipeline decisions can be reused as deterministic assets, while unknown failures stop or escalate. The published case favors serial operation under its current constraints and does not claim that greater concurrency is inherently better. This is useful selection evidence: an appropriate framework should record a justified decision to omit Parallel or open-ended reasoning.

Our experiment implements only the acceptance and recovery boundary. A file-backed service surrogate returns status, content type, and bytes, including generated PNG images. The consumer decodes the image and checks whether it is uniform. The independent task oracle knows which image belongs to the intended source. No GeoServer, GDAL, actual geospatial dataset, network stack, or browser is executed. This restricted environment allows repeatable tests of false-success detection, known-signature repair, and refusal of unknown errors; it cannot establish geographic accuracy, production latency, or service scalability.

\subsection{Shared artifacts: evidence and event lifecycle}
Fan's design notes examine how cooperating agents exchange artifact references, read content on demand, process repeated events, and determine when stored content can be reclaimed~\cite{fanproposal}. The proposal connects content checks with scope and identity checks, and connects delivery with locks, state transitions, and acknowledgments. Its contribution to this study is a concrete collaboration problem and a set of candidate mechanisms. Deployment and performance claims are not inferred from the proposal's use of industrial terminology.

The reconstruction uses two consumers, a real artifact file, and SQLite effects and receipts. One factor validates scope, version, and a digest against a trusted manifest. The other records an effect and its durable receipt atomically, recognizes repeated delivery, and permits collection after two distinct completion receipts. These mechanisms address different failure surfaces. A correct digest does not prevent repeated side effects; a receipt does not establish that the bytes belonged to the intended task.

The implementation deliberately uses a single transactional store~\cite{sqlite}. It does not claim exactly-once execution across external services, prove a distributed-lock protocol, or simulate concurrent workers and network partitions. Fan's proposed lock, identity, streaming, and storage-tier choices therefore remain candidates for implementation-specific replication. The present result concerns local crash/replay and artifact lifecycle semantics, not large-file throughput or production infrastructure.

\subsection{What transfers across cases}
Across the industrial accounts, a local success signal fails to establish an end-to-end property. A valid identifier may address the wrong record, a registered layer may be unusable, and a delivered event may not represent a new operation. The cross-case inference is that selection should begin with the property that must survive a boundary: identity, acceptance evidence, or effect uniqueness. The mechanism and topology then have a concrete obligation to satisfy.

The SysML research case contributed by Liangding Yuan provides a related transfer setting: a plausible diagram can hide invalid repository ownership or unresolved relationship endpoints~\cite{yuancase}. We use it as contextual triangulation and as a possible future held-out design task. We do not add a fourth runtime experiment or count its screenshots as repeated trials. Three principal scenarios already permit the study to examine distinct state, acceptance, and collaboration boundaries.

\section{Selecting a Composition from a Failure}
\label{sec:method}
\subsection{Coordinates, contracts, and graph structure}
Let a pattern instance be
\begin{equation}
 p=(f,t,I,O,V,K),
\end{equation}
where $f$ is its primary cognitive responsibility, $t$ its primary execution topology, $I$ and $O$ its input and output contracts, $V$ its verification obligation, and $K$ its resource and authority constraints. A composition is a graph $\mathcal{A}=(P,E)$ over pattern instances with typed handoffs. Along an edge $(p_i,p_j)$, the receiving contract must accept the actual upstream artifact, including its identity, version, scope, and readiness state. Shared state and authority need explicit owners even when the graph is drawn as a simple chain.

The original paper describes seven functions and six topologies~\cite{huangzhou}. The ADPS catalog snapshot inspected for this study contains 27 cell-bound patterns and three cross-cutting engineering planes: observability, evaluation and validation, and security and identity~\cite{catalog}. Governance remains a cognitive-function row; the cross-cutting planes do not replace it or create additional topology columns. Pattern identifiers and catalog versions belong in the experiment record because the catalog has evolved since the earlier paper's 28-pattern version.

Orthogonality is a representational claim: functional responsibility and control structure answer distinguishable questions. It does not imply statistical independence of their effects. In the map example, verification must expose an error before a recovery loop can respond to it. Interaction between the two components is an expected consequence of composition, not a contradiction of the two-axis representation.

\subsection{Catalog admission and retirement}
\label{sec:admission}
Two questions are easy to conflate. Whether a pattern belongs in the catalog is a question about a shared vocabulary. Whether adding it improves a particular system is a question about one engineering budget. Only the second is settled by the controlled comparisons reported below. The first is not an empirical proposition of that kind: no ablation establishes that a pattern is worth naming, any more than a trial establishes that a term belongs in a glossary. Design patterns belong to what Simon called the sciences of the artificial~\cite{simon}; their warrant is generative rather than predictive.

We therefore state the admission criteria explicitly, because a catalog without them drifts. A cell is admitted, and retained, only while all four hold.

\begin{enumerate}
\item \textbf{Independent recurrence.} The structure has been implemented separately in systems that did not copy one another. The evidence is source code and incident records, not practitioner endorsement. Testimonial support is the weakest form available here, and it is the form that self-help literature also enjoys.
\item \textbf{Separability.} The decision the pattern isolates is not already isolated by an existing cell. A candidate that cuts no new decision is merged into the cell that already covers it. A candidate that cuts a real decision spanning several functional rows is promoted to a cross-cutting engineering plane instead of occupying one cell.
\item \textbf{Describable failure on omission.} A team unaware of the pattern falls into a specific, reproducible failure, not a diffuse shortfall in quality.
\item \textbf{Effect on design action.} A practising designer recognizes the structure on sight and subsequently changes what they build. A pattern that is agreed with but never acted on fails this criterion.
\end{enumerate}

Retirement is the same test applied later. A worked example occurred during this study. Observability had occupied the governance row as cell G4 under an orchestration topology. It fails the separability criterion: the decision it isolates recurs across main agents, sub-agents, tool invocations, and human takeover, so a single governance cell understates its scope without isolating anything the other governance cells leave open. It was therefore removed from the matrix on 11 September 2026 and promoted to the cross-cutting plane X1, alongside evaluation and validation (X2) and security and identity (X3). The matrix moved from 28 cells to 27; the cross-cutting planes add no topology column and occupy no cell. This is the change behind the catalog version difference noted above, and it is also the first demonstration that the criteria can remove an entry rather than only admit one. Earlier publications that cite the governance coordinate remain correct under the catalog version they name.

These criteria govern membership only. They are deliberately not evidence that a composition helps a given system, and they are not interchangeable with the ablations reported later.

\subsection{An evidence-driven selection procedure}
The procedure begins with the smallest externally verifiable business operation and proceeds through six decisions:
\begin{enumerate}
\item Fix the target state, permissible side effects, and evidence required for acceptance.
\item Execute the simplest defensible baseline and identify the first observable divergence.
\item Assign the missing responsibility and select its primary functional coordinate.
\item Select the least complex control structure that preserves dependencies, stopping conditions, and authority.
\item Specify handoff contracts and a counterexample for every claimed benefit of the composition.
\item Compare the baseline, individual components, and composition; retain the observed trade-offs and a condition for revisiting the decision.
\end{enumerate}
For the business case, the divergence is a misbound identifier or an out-of-order effect. The functional response binds tool arguments to authoritative records. The execution response requires a snapshot before applying the update and records completion with the effect. For map publication, the divergence lies after the apparent success of the publishing call, so further reasoning before that call does not supply the missing consumer evidence. Selection moves the observation point and then adds only the recovery routes supported by the observations.

Each decision retains an alternative. If normal inputs follow a stable certified pipeline, open-ended runtime planning may add no measurable value. If failure cannot be distinguished by a trustworthy check, a more persistent retry loop can repeatedly execute the wrong operation. A selection record should therefore include why a pattern was rejected or removed, as well as why one was adopted.

\subsection{Pattern Evaluation Contract}
The comparison contract records the task and source version, proposed action, model or proposal generator, tool implementation, environment and fault schedule, enabled mechanisms, call graph, stopping rule, observable artifacts, oracle, and reporting rule. The contract distinguishes the system's \emph{runtime verifier} from the study's \emph{external oracle}. Removing a runtime check is a treatment; removing the corresponding test oracle would change the outcome definition and invalidate the comparison.

\begin{figure}[t]
\centering
\begin{tikzpicture}[>=Stealth, every node/.style={font=\small,align=center},
box/.style={draw=ink,minimum height=1.1cm,text width=3.0cm,inner sep=5pt}]
\node[box] (case) {Industrial failure\\and source artifact};
\node[box,right=8mm of case] (select) {Function and topology\\selection contract};
\node[box,right=8mm of select] (run) {Executable alternatives\\and identical faults};
\node[box,below=9mm of run] (oracle) {External artifact oracle\\and false-success check};
\node[box,left=8mm of oracle] (decision) {Measured trade-off\\and revision condition};
\draw[->] (case)--(select);\draw[->] (select)--(run);
\draw[->] (run)--(oracle);\draw[->] (oracle)--(decision);
\draw[->] (decision)--(select);
\end{tikzpicture}
\caption{Evidence links a reported failure to an executable treatment and a decision that can be revisited. The external oracle remains fixed when the runtime verifier is ablated.}
\label{fig:evidence}
\end{figure}
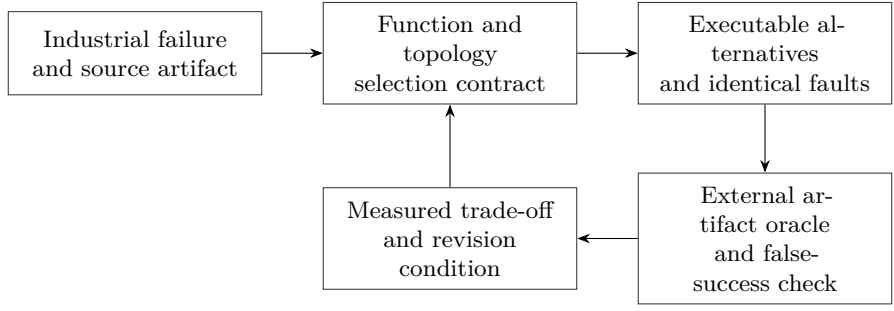

The workshop synthesis supplies six measurement surfaces: outcome, trajectory, control, recovery, economy, and evolution. These are reporting surfaces, not a composite leaderboard score. A current experiment may measure only part of them. The reference study observes final state or bytes, event order, simple admission checks, bounded recovery, and instrumented operation counts. It records reproducible versions for later evolution but does not measure longitudinal maintenance benefits or human review time.

\section{Reference-System Experiment}
\label{sec:experiment}
\subsection{Treatments and controls}
We implemented three local reference systems and varied two binary treatment bundles: function-specific contracts $F$ and execution controls $T$. The primary arms are $B_{00}$, $B_{10}$, $B_{01}$, and $B_{11}$. Table~\ref{tab:treatments} defines their executable differences. These bundles operationalize responsibilities and control boundaries; they do not exhaust a cognitive category or isolate one universal topology effect. For example, durable control introduces persistent state as well as ordering.

\begin{table}[t]
\centering\small
\caption{Two treatment bundles in each reference system. All four combinations use the same environment constructor and task oracle.}
\label{tab:treatments}
\begin{tabularx}{\textwidth}{@{}p{2cm}YYY@{}}
\toprule
System & $F$: function-specific contract & $T$: execution control & Example pattern coordinates\\
\midrule
Business & Bind the action to a trusted record; check task authority & Fix dependency order; commit effect and completion together; resume from checkpoints & Action/Memory responsibilities; Chain with durable orchestration and Governance admission\\
Artifacts & Validate allowed scope, version, and bytes against a trusted digest & Atomic effect/receipt; deduplicated consumption and completion; collection after all receipts & Collaboration handoff; Memory retention; bounded event execution\\
Maps & Decode consumer response; inspect status, type, body, and image uniformity & Route known observed errors to repair; bound retries; stop on unknown errors & Perception and verification; Reasoning/Reflection with Route and Loop\\
\bottomrule
\end{tabularx}
\end{table}

A retry-only control increases the business and artifact retry ceiling from three to six attempts; for maps it retries a failed publication up to six times without consumer verification. This tests a limited alternative explanation based on repeated attempts. It is not an equal-token or equal-latency control. No model tokens are consumed, and different architectures execute different operations.

A same-code relabeling control, named \texttt{standard} in the artifact, enables exactly the same mechanisms as $B_{11}$. It checks that outcomes and operation counts do not depend on an ADPS label. This is a deterministic identity check, not an independently designed expert baseline or evidence of superiority over established engineering practice. Both implementations use known controls such as validation, state machines, and transactions.

\subsection{Scenarios, execution, and independence}
Each system has 12 scenario families: normal operation, component-specific faults, compound faults, and a boundary case with wrong trusted evidence. The complete inventory appears in Appendix~\ref{app:scenarios}. Ten deterministic seed variants change identifiers, values, artifact bytes, or map content within each family. Six arms yield $3\times12\times10\times6=2160$ runs. All variants produced the same categorical outcome within each arm--family cell. We therefore report family counts as the principal result; the variants establish only input-level stability within this generator.

The suite is exploratory and constructed from the case mechanisms. A one-variant development pass preceded the reported run and revealed an inconsistency in the completion metric: a correct artifact with an unsuccessful run report had been counted as a correct resolution. The definition was corrected before the reported run. The final source hash, scenario inventory, arm definitions, and shuffled schedule seed were written to a manifest before that run began. This is a local execution freeze, not a public preregistration or an independently hidden test set. No scenario was removed after seeing the final results.

The reported run used Python 3.9.6, SQLite 3.51.0, and Pillow 11.3.0 on macOS. SQLite transactions produce actual effects and completion records. Injected exceptions before or after a committed effect close and reopen the database connection, representing application-level restart and replay. They do not test process termination during a storage write, power loss, distributed transactions, or concurrent execution. Map and artifact tests read actual bytes from local files. No production connection, paid API, current language-model inference, or external user action is involved.

Action proposals are generated deterministically, including explicit wrong-identifier and wrong-order perturbations. The causal comparison is therefore conditional on those proposals and faults. It measures whether the surrounding architecture prevents or recovers the specified failure. It does not measure the probability that a language model produces the proposal, nor an end-to-end success rate in a natural task distribution.

\subsection{Outcomes and artifact audit}
Let $C_i$ indicate that a run claims completion, $O_i$ indicate that its final artifact satisfies the external oracle, and $R_i$ indicate an explicitly permitted, correctly executed refusal or bounded stop. We report
\begin{align}
\text{accepted completion}_i &= C_i O_i,\\
\text{false success}_i &= C_i(1-O_i),\\
\text{correct resolution}_i &= (C_i O_i)\lor R_i.
\end{align}
A refusal never counts as a completed business operation. In particular, unauthorized updates and invalid artifacts require refusal. Persistent or unknown map errors permit a bounded stop. These resolution policies are part of the scenario specification; they are not selected retrospectively to favor an arm. Reporting completion and refusal separately prevents a reject-everything system from appearing useful.

Business oracles inspect the intended record, unchanged unrelated record, effect count, and snapshot value. Artifact oracles require one effect for each of two consumers using the intended digest. The map oracle requires the generated image associated with the intended source. A meaningful but wrong map can therefore fool the runtime's image check while failing the experiment's oracle. This is an intentional difference between limited observation and task truth.

A separate audit program imports no runtime or oracle functions. It opens the final databases, reads the artifact bytes, verifies recorded refusal events, and reconstructs all four categorical outcomes. All 2160 records agree with this readback. This is independent code-path verification, not independent human annotation: both programs were developed within the same AI-assisted research workflow. The archive retains the distinction.

\subsection{Analysis}
We report fixed-suite counts, differences, per-family outcomes, and operation counts. The 36 families are purposively selected rather than sampled from a population, and the ten variants are not independent industrial cases. We accordingly do not calculate population $p$-values or present a narrow confidence interval based on 2160 apparently independent observations. For an outcome $Y$, the descriptive interaction contrast is
\begin{equation}
 I_Y=\bar Y_{11}-\bar Y_{10}-\bar Y_{01}+\bar Y_{00},
 \label{eq:interaction}
\end{equation}
with equal weighting of the 12 families in the relevant system. The magnitude depends on this suite and weighting. A positive contrast for completion indicates complementarity under these scenarios; it does not establish a universal law of composition.

\section{Reference-System Results}
\label{sec:results}
\begin{table}[t]
\centering\small
\caption{Main results as scenario families out of 12 per system. Every categorical cell is unchanged across the ten input variants. Each cell is accepted completion / safe refusal or stop / false success. Failed runs without a false claim account for the remainder.}
\label{tab:main}
\begin{tabular}{@{}lrrrr@{}}
\toprule
System & $B_{00}$ & $B_{10}$ & $B_{01}$ & $B_{11}$\\
\midrule
Business execution & 2 / 0 / 8 & 4 / 2 / 6 & 6 / 0 / 4 & 9 / 2 / 1\\
Shared artifacts & 2 / 0 / 5 & 2 / 4 / 2 & 7 / 0 / 5 & 7 / 4 / 1\\
Map publication & 1 / 0 / 8 & 1 / 3 / 1 & 2 / 0 / 10 & 8 / 3 / 1\\
\bottomrule
\end{tabular}
\end{table}

\subsection{Correct information and durable execution address different faults}
In the business system, binding actions to authoritative records resolves proposal drift and wrong-but-valid identifiers. It does not preserve dependency order or prevent an update from being applied again after an interrupted return. Durable execution improves accepted completion from two to six families while leaving incorrect or unauthorized record selection unresolved. Composition completes nine families and safely refuses two unauthorized families. The remaining family supplies the wrong record through the trusted input itself.

The associated traces explain the compound effect. Without durable completion, interruption after an update commits causes the next attempt to apply the increment again. With durable completion, the effect and completed-step record commit together, and a reopened connection observes that the step has already run. Yet this mechanism remains faithfully wrong when the selected identifier points to the wrong valid record. Combining correct binding with durable execution is necessary for the drift-and-replay scenario. The completion interaction is $1/12$, or 8.3 percentage points, on the fixed suite.

\subsection{Artifact integrity and delivery reliability contribute differently}
Artifact contracts alone reject corrupted bytes, stale versions, and unauthorized reads, but accepted completion remains two families. Durable receipts and unique acknowledgments increase accepted completion to seven families. Their benefit is observable in concrete file and database state: a duplicate completion acknowledgment no longer deletes content before the second consumer reads it, and a repeated event after collection can be recognized from its committed receipt without rereading the deleted file.

Composition retains seven completed families and adds four correct refusals. The interaction for accepted completion is zero: its gains are additive under this outcome definition, not evidence of universal synergy. The combination further reduces false-success families from two under contracts alone to one because replay and premature collection interact with the reporting path. A reconstructed event stream is sufficient for this local result, but a deployed message broker, lease expiry, delayed consumer, and externally committed side effect remain separate validation targets.

\subsection{Recovery can amplify false success}
The map system produces the largest interaction. Consumer verification alone detects most unacceptable outputs but cannot repair them. Execution control alone retries a failed publication and routes visible publication errors; it cannot react to an error that the runtime never observes. This arm increases false-success families from eight to ten because two compound scenarios pass the transient publishing failure and then stop at an invalid consumer result that is never inspected.

The composed system uses observed signatures to repair known layer, blank-image, or metatile conditions and rechecks the consumer result. It completes eight families and stops on three persistent or unknown errors. Its completion interaction is $6/12=0.50$. The result illustrates a selection constraint: adding recovery increases the importance of an adequate acceptance test. A loop can make an insufficient completion rule more effective at producing an incorrect success report.

\begin{figure}[t]
\centering
\begin{tikzpicture}
\begin{axis}[width=.94\textwidth,height=5.4cm,ybar,bar width=12pt,
ymin=0,ymax=12,ytick={0,2,4,6,8,10,12},ylabel={Correctly resolved families / 12},
symbolic x coords={Business,Artifacts,Maps},xtick=data,
enlarge x limits=.22,legend columns=4,legend style={at={(.5,1.08)},anchor=south,draw=none},
ymajorgrids=true,grid style={gray!20},tick label style={font=\small},legend cell align=left]
\addplot[fill=ink!65,draw=ink] coordinates {(Business,2) (Artifacts,2) (Maps,1)};
\addplot[fill=ochre!65,draw=ochre] coordinates {(Business,6) (Artifacts,6) (Maps,4)};
\addplot[fill=brick!65,draw=brick] coordinates {(Business,6) (Artifacts,7) (Maps,2)};
\addplot[fill=teal!75,draw=teal] coordinates {(Business,11) (Artifacts,11) (Maps,11)};
\legend{$B_{00}$,$B_{10}$,$B_{01}$,$B_{11}$}
\end{axis}
\end{tikzpicture}
\caption{Correct resolution includes separately specified refusals and bounded stops. It is not a task-completion rate. The fixed, deliberately fault-heavy suite gives equal weight to each family.}
\end{figure}
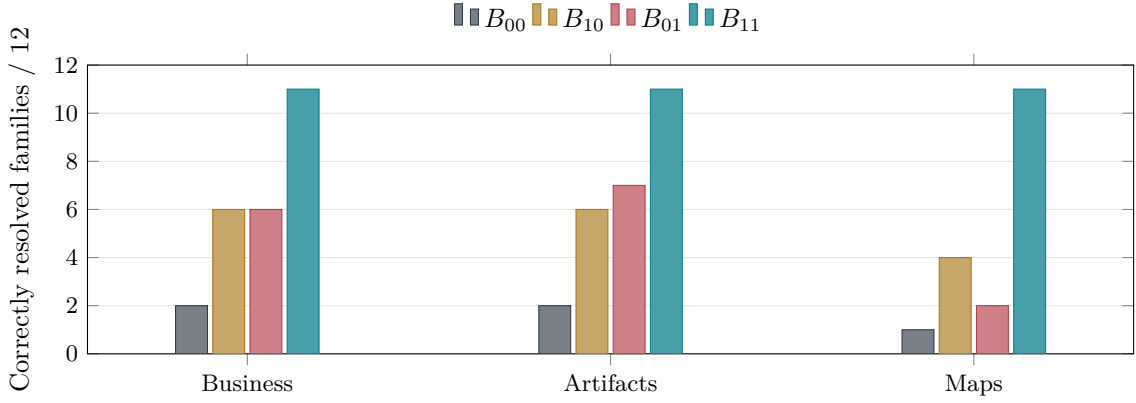

\subsection{Additional attempts, relabeling, and clean-path overhead}
The retry-only arm changes no categorical result in the business or artifact suites. In maps it completes two families but falsely claims success in ten, matching the control-only arm's categorical result. More attempts can address an exposed transient failure; they do not repair an unobserved consumer failure or bind an identifier to its source. These findings apply to the tested retry schedules, not all adaptive policies.

All 360 paired inputs to $B_{11}$ and the same-code control produce identical categorical results and instrumented operation counts. That equality is expected and serves as an implementation check. The experiment gives no evidence that attaching a pattern label adds runtime effectiveness, or that an experienced engineer could not construct the same architecture without the framework.

Clean cases also expose a cost. In business execution, instrumented operations increase from two to five; in artifact exchange, from six to twelve; and in map publication, from one to three. These counts include checks and state reads, not just external API calls, and should not be converted into monetary or token costs. Host timings are archived as execution metadata, but the small local workload does not justify a production latency claim. The extra controls must be justified by failure exposure and consequence, rather than by assuming that every request needs the full composition.

\subsection{All three compositions retain a boundary failure}
The composed business system uses an incorrect authoritative record correctly. The composed artifact system validates altered content against a manifest that has already been changed to endorse that content. The composed map system accepts a nonuniform, decodable image that belongs to the wrong source. Each fails one boundary family across all ten variants.

These failures limit the architectural guarantee. Provenance establishes where a value came from; it does not establish that the source selected the intended object. Digest validation establishes agreement with a trusted expected digest; it does not establish the truth of an untrusted or incorrect manifest. Image presence establishes renderability; it does not establish geographic or semantic correctness. Stronger acceptance may require an independent source, domain-specific invariants, or human judgment. No additional retry budget supplies that missing evidence.

\section{Historical Model Pilot: Auditing the Evaluator}
\label{sec:legacy}
The reference systems test deterministic execution boundaries. To retain contact with model-generated task artifacts, we also reanalyzed the frozen RegTech pilot from the earlier manuscript. It extracts obligations from a bundle based on the Monetary Authority of Singapore's Fair Dealing guidelines~\cite{mas}. The six arms comprise a single pass, critic and repair, iterative self-refinement, candidate-guided review, a union queue, and consensus refinement. Each has one recorded run with model alias \texttt{gpt-4}, temperature 0.1, and a 12-item reviewed checklist.

\begin{table}[t]
\centering\small
\caption{Recomputed historical pilot scores under its original matcher. These are lexical/structural matching scores, not independently adjudicated semantic coverage.}
\label{tab:legacy}
\begin{tabular}{@{}lrrrrr@{}}
\toprule
Arm & Predicted items & Matches & Precision & Recall & F1\\
\midrule
Single pass & 16 & 6 & .3750 & .5000 & .4286\\
Critic and repair & 16 & 7 & .4375 & .5833 & .5000\\
Self-refinement & 16 & 7 & .4375 & .5833 & .5000\\
Candidate-guided review & 16 & 9 & .5625 & .7500 & .6429\\
Union queue & 29 & 10 & .3448 & .8333 & .4878\\
Consensus refinement & 16 & 8 & .5000 & .6667 & .5714\\
\bottomrule
\end{tabular}
\end{table}

The matcher combines ROUGE-L, token Jaccard overlap, source-section agreement, category agreement, and predicate-type agreement with weights .35, .20, .20, .15, and .10. It greedily matches prediction--gold pairs one-to-one at threshold .36. All seven stored summary fields reproduce exactly for every arm. However, removing source-section agreement and renormalizing the remaining weights reverses the comparison: candidate-guided review falls from F1 .6429 to .2857, while single pass remains .4286. The scorer therefore rewards a structural property that the candidate-guided arm is designed to preserve.

This sensitivity does not establish that the alternative scorer is correct; its threshold is not recalibrated after changing the score distribution. It establishes that the ranking is not robust to a plausible measurement change. Source traceability remains useful, while improved semantic obligation coverage requires separate annotation. The historical paired diagnostic reports four checklist items matched only by candidate review and one only by the baseline, giving an exact two-sided value of .375. A bootstrap over those 12 checklist items yields a recall-difference interval of $[-.083,.583]$. Both calculations are conditional on this one document and matcher; correlated obligations cannot support a document-population inference.

Additional limitations are material. The candidate extractor and gold checklist originated in the same authoring workstream, token usage was not recorded, the historical alias is not an immutable model snapshot, and the arms change multiple components and budgets. The union queue reuses upstream outputs, so its assembly time is not end-to-end latency. Extra self-refinement matches the critic arm's coverage while taking longer and reducing schema compliance. These results motivate evaluator ablation, separate reporting of review queues and final outputs, and explicit budget accounting in subsequent model experiments.

\section{From Current Evidence to Continuing Industrial Validation}
\label{sec:evolution}
The community can sustain this study by contributing reproducible design decisions rather than an ever-growing collection of favorable anecdotes. A contribution should connect a bounded task, an encountered failure, the previous and revised architecture, a replayable input or trace, an external acceptance rule, and a known applicability limit. The artifact records which material is public, which measurements remain inside the organization, and who can verify the system-specific interpretation. It also records whether the framework preceded the design decision or was applied retrospectively. A retrospective mapping cannot establish that the framework caused adoption.

The current cases suggest a practical division of work. Liang's system can contribute redacted dependency and parameter-provenance traces with readback states. Xiong's system can contribute real publication requests, output images, pipeline versions, and failure-triggered recertification records. Fan's proposal can advance to an executable consumer and broker integration, with tests around delivery, receipt persistence, authority changes, and artifact retention. These contributions would replace specific surrogate assumptions with observed platform behavior. They are proposed next measurements, not work already performed by the listed contributors for the new benchmark.

An evolving benchmark needs three distinct collections: a public regression set, a sequestered evaluation set owned by a reviewer outside the implementation team, and incoming field incidents awaiting adjudication. Once a hidden case guides a repair, it becomes development or regression material; a later evaluation needs fresh withheld cases. Every result should retain a tuple of catalog, case-set, runtime, model, and oracle versions. Changing an oracle creates a new measurement series and requires rescoring a fixed set of outputs so that an apparent architecture improvement is not a grader change.

For deployments, begin with replay or shadow execution that cannot commit live side effects, then compare candidate and incumbent versions under the same acceptance policy. Record attempted operations, verified effects, false successes, authorized refusals, intervention time, and regressions. A migration from model-driven execution to a certified deterministic pipeline may be a successful outcome of framework use. The objective is a better justified architecture for the task, including removal of unnecessary patterns.

\section{Testing Whether the Framework Helps Engineers}
\label{sec:human}
A human design study is the smallest additional study that directly answers the framework-effectiveness question. It should compare three conditions: a conventional engineering checklist, the same pattern descriptions in a flat catalog, and the same descriptions organized by the two axes with the selection procedure. All conditions receive the same business requirements, examples, tool interfaces, preparation time, and implementation budget. The flat-catalog comparison isolates the value of organization and selection guidance from the value of merely giving engineers more information.

Use a parallel-group assignment stratified by experience and prior ADPS exposure. A crossover on the same design method risks irreversible learning carryover. Each engineer works on matched unfamiliar tasks, such as a repository write-back scenario adapted with its contributor and a new stateful service task. Contributors should not be evaluated on their own published cases. Reviewers assess de-identified implementation artifacts, and neither the rubric nor the executable acceptance tests award points for naming ADPS patterns. A single short feasibility cohort, for example 24 engineers with eight per condition, can estimate recruitment, task duration, and rating variance. It is not automatically a powered confirmatory sample. A larger sample must be justified from the smallest practically meaningful effect and clustering by engineer and task.

The primary outcome should be externally verified requirement satisfaction within the fixed engineering budget. Secondary outcomes include critical defects, time to a first accepted implementation, justified omission of unnecessary components, and effort to handle a subsequent change request. Design diagrams and self-reported confidence are diagnostic outcomes. For the main contrast, report the difference between the two-axis condition and the flat catalog, with uncertainty at the engineer level; comparisons to the generic checklist provide context. Prespecify exclusions, missing deliveries, and the treatment of unfinished systems. This study has not been executed in the present work, and its result must remain open.

\section{Threats to Validity}
\textbf{Construct validity.} The reference outcomes measure selected architectural obligations, not the full quality of an agent. The function and control treatments are mechanism bundles. Correct resolution includes specified safe refusals, so it must be read alongside accepted completion and false success. Instrumented operations have unequal cost, and an image-validity check is not a geospatial correctness test.

\textbf{Internal validity.} The same source corpus informed both the treatment design and the challenge suite. The injected faults were chosen because the controls should address them. This supports mechanism testing but risks favorable benchmark construction. The same dependency limits what the results say about selection: the mechanisms were already known from the cases before the procedure was applied, so the experiments show that the selected mechanisms work, not that the procedure would have found them. Evaluating the procedure itself requires the design study of Section~\ref{sec:human}. Compound and wrong-evidence cases provide useful counterexamples without eliminating that bias. Both implementation and audit code were AI-assisted and need contributor and co-author review. The final execution freeze followed a development pass and must not be described as preregistration.

\textbf{External validity.} Two industrial accounts, one contributor proposal, and three small local systems cannot represent all agent architectures. The artifact payloads are small, map outputs are synthetic, and business state is local. There are no concurrent workers, real broker failures, GeoServer integration, current-model runs, or independently gathered production outcome rates in the new experiment. The legacy pilot adds model-produced artifacts but only from one document and one run per arm.

\textbf{Conclusion validity.} Ten variants per family do not create ten independent tests of an industrial hypothesis. Equal family weighting emphasizes deliberately selected failures and should not be used to forecast production rates. All main contrasts are descriptive. The same-code control validates identity, not comparative performance against an independent expert implementation. Strong reference-system performance cannot substitute for the unperformed human design study.

\textbf{Case and researcher relationships.} The framework's originator, community participants, and case contributors are involved in the research. The cases are purposively selected and may favor problems the catalog already expresses well. Source attribution preserves the record but does not eliminate allegiance bias. External reviewers, systems not designed by the authors, and fresh field cases are needed for stronger replication. Planned contributor verification is not reported as completed member checking.

\section{AI Assistance and Research Responsibility}
Generative AI contributed substantially to preparation of this manuscript and its research artifacts. OpenAI Codex assisted with searching and synthesizing the supplied materials, developing the experimental design, implementing the reference systems and audit programs, interpreting results, and drafting substantial portions of the English text in the September 2026 revision. Earlier working drafts also used AI assistance. Exact model snapshot identifiers were not consistently retained, so tool-level attribution should not be read as a reproducible model specification. This involvement extends beyond proofreading or translation.

The underlying two-axis framework, industrial accounts, contributor proposals, and workshop discussions are attributed to their human sources. The new numerical results come from executing the disclosed programs and reading their output artifacts. AI-generated prose is not an observation, and AI review is not independent human adjudication. No new participant responses, production telemetry, or current-model benchmark outputs were synthesized and presented as collected data. Responsibility for the research claims, citations, permission to use contributed material, and the final submitted text rests with the named human authors. AI tools are not listed as authors. Co-author approval of this manuscript is a separate human step and is not implied by its automated preparation.

\section{Conclusion}
The industrial material identifies engineering needs that a final-answer benchmark alone would miss: preserving the identity of business state, verifying a result from its consumer, and connecting event delivery to durable effects. A two-axis selection procedure makes the functional responsibility and execution condition explicit, while a composition contract turns the proposed benefit into a counterexample and an executable comparison.

The reference experiments show conditional benefits of composing these mechanisms, additive rather than synergistic behavior in one setting, a false-success regression from recovery without verification, and failures when trusted evidence is wrong. The historical model pilot adds a different warning: reproduced scores can still depend on the evaluator. These results justify particular architecture choices within stated boundaries. Catalog membership rests on the stated admission criteria rather than on these experiments, and the promotion of observability out of the matrix shows those criteria removing an entry. A direct demonstration that the framework improves engineers' designs requires the separate controlled study described here. The continuing research asset is the connection between a contributor's observed problem, a traceable design decision, a reproducible test, and the evidence that would cause that decision to change.

\appendix
\section{Complete Scenario Inventory}
\label{app:scenarios}
The families below were included in the reported run. Names denote environment conditions, not information exposed to the runtime policy. Ten seeds, 1000--1009, instantiate each family. Their categorical results were constant within each arm--family cell.

\begin{table}[h]
\centering\footnotesize
\caption{Twelve families per reference system. The final row is an explicit boundary test of the composed implementation.}
\begin{tabularx}{\textwidth}{@{}rYYY@{}}
\toprule
& Business execution & Shared artifacts & Map publication\\
\midrule
1 & Clean update & Clean two-consumer delivery & Clean publication\\
2 & Altered proposal identifier & Duplicate delivery & HTTP 200 with XML error\\
3 & Wrong valid identifier & Crash before effect & Blank rendered image\\
4 & Reordered steps & Crash after effect & Zero metatile error\\
5 & Crash before effect & Corrupt artifact bytes & Transient publication failure\\
6 & Crash after effect & Stale artifact version & Transient consumer failure\\
7 & Duplicate delivery & Unauthorized consumer & Missing source CRS\\
8 & Unauthorized update & Duplicate acknowledgment & Persistent service error\\
9 & Identifier drift and replay & Early collection event & XML error and publication restart\\
10 & Reordered steps and replay & Corruption and replay & Blank image and publication restart\\
11 & Unauthorized update and replay & Replay after collection & Image type with nonimage bytes\\
12 & Wrong authoritative record & Wrong trusted manifest & Plausible image from wrong source\\
\bottomrule
\end{tabularx}
\end{table}

The reference source hash for the reported execution is\par
\noindent\texttt{344145a29b043809a5619c49296621b76b3}\par
\noindent\texttt{090d6058beead47162dc6eba67144}.
The two lines concatenate to one SHA-256 digest. The manifest records execution time, dependency versions, arms, and schedule seed. A fresh run must use a new output directory to preserve earlier artifacts.

\section{Reproduction and Artifact Status}
The reference-study replication ZIP contains \texttt{benchmark.py}, \texttt{audit.py}, a dependency specification, the frozen manifest, per-run results, per-family results, compressed event traces, and audit reports. The first script requires Python and Pillow and calls no external service. The second reconstructs results from the generated databases and files without importing runtime scoring functions. The local project additionally retains \texttt{audit\_legacy.py} and its historical audit output. That script reuses the historical scorer's computation functions through structured Python AST extraction, avoiding its plotting dependencies and leaving the old files unchanged. Historical source documents and model outputs are not included in the reference-study ZIP; it reproduces the new reference experiments, not the historical model pilot.

\begin{verbatim}
python benchmark.py --output named-reference-run --variants 10
python audit.py named-reference-run
\end{verbatim}
The same command regenerates the databases and response files needed for the audit. Timing fields may differ and are excluded from deterministic comparisons. Synthetic traces contain no participant responses or production identifiers. Public release of the complete artifact requires the contributors' permissions and a stable repository or archive location; no uncreated public dataset or DOI is claimed here. Public case reports are cited below, and the unpublished collaboration notes retain attribution without exposing private contact details.

\end{document}